\documentclass[11pt]{revtex4}
\usepackage{amssymb,epsf}
\usepackage{latexsym}

\begin{document}

\title{Charged rotating black string in gravitating nonlinear electromagnetic
fields}
\author{S. H. Hendi\footnote{hendi@shirazu.ac.ir} and A. Sheykhi\footnote{asheykhi@shirazu.ac.ir}}
\affiliation{Physics Department and
Biruni Observatory, College of Sciences, Shiraz
University, Shiraz 71454, Iran \\
Research Institute for Astrophysics and Astronomy of Maragha (RIAAM), P.O.
Box 55134-441, Maragha, Iran}

\begin{abstract}
We obtain a new solution of rotating black string coupled to a
nonlinear electromagnetic field in the background of anti-de
Sitter spaces. We consider two types of nonlinear electromagnetic
Lagrangians, namely, logarithmic and exponential forms. We
investigate the geometric effects of nonlinearity parameter and
find that for large $r$, these solutions recover the rotating
black string solutions of Einstein-Maxwell theory. We calculate
the conserved and thermodynamic quantities of the rotating black
string. We also analyze thermodynamics of the spacetime and verify
the validity of the first law of thermodynamics for the obtained
solutions.
\end{abstract}

\maketitle


\section{Introduction}

The theory of nonlinear electrodynamics was first introduced in
$1934$ by Born and Infeld for the purpose of solving various
problems of divergence appearing in the Maxwell theory \cite{BI}.
In recent years, the study of nonlinear electrodynamics has got a
new impetus. Strong motivation comes from developments in
string/M-theory, which is a promising approach to quantum gravity
\cite{BIString}. It has been shown that the Born-Infeld (BI)
theory naturally arises in the low energy limit of the open string
theory \cite{Frad,Cal}. Another motivation originates from the
fact that most physical systems in the nature, including the field
equations of the gravitational systems, are intrinsically
nonlinear. The nonlinear BI electromagnetic theory was designed to
regulate the self-energy of a point-like charge \cite{BI}. Various
aspects of black hole solutions coupled to nonlinear BI gauge
field has been studied. Exact solutions of the Einstein BI
theory with or without cosmological constant have been constructed in \cite%
{Fern,Tamaki,Dey,Cai1,MHD}. In the scalar-tensor theories of gravity, black
object solutions coupled to a Born-Infeld nonlinear electrodynamics have
also been studied widely in the literature \cite{Tam,SheyBI}.

However, BI theory is not the only nonlinear electrodynamics theory which
can remove the divergence of the electric field at $r = 0$. In particular,
in recent years, other types of nonlinear electrodynamics in the context of
gravitational field have been introduced. In \cite{Soleng} exact solution
for a static spherically symmetric field outside a charged point particle is
found in a nonlinear $U(1)$ gauge theory with a logarithmic Lagrangian.
While this particular theory appears to have no direct relation to
superstring theory, it serves as a toy model illustrating that certain
nonlinear field theories can produce particle-like solutions which can
realize the limiting curvature hypothesis also for gauge fields \cite{Soleng}%
. In addition to BI and logarithmic types for nonlinear gauge
fields, very recently one of the present authors proposed an
exponential form of nonlinear electromagnetic Lagrangian
\cite{HendiJHEP}. Although a logarithmic form of the
electrodynamic Lagrangian, like BI electrodynamics, removes
divergences in the electric field, the exponential form of
nonlinear electromagnetic Lagrangian does not cancel the
divergency of the electric field at $r = 0$, however, its
singularity is much weaker than in the Einstein-Maxwell theory.
Other studies on the gravitational systems coupled to nonlinear
electrodynamics gauge fields have been carried out in \cite%
{PMIpaper,Oliveira,Soleng2}.

The extension of the Maxwell field to the nonlinear
electromagnetic gauge field provides powerful tools for
investigation of black object solutions. In the present work, we
would like to turn the investigations on the nonlinear
electrodynamics to the rotating black string solutions with one
rotation parameter in the anti-de Sitter (AdS) background. We will
consider the four dimensional action of Einstein gravity with two
kinds of BI type nonlinear electromagnetic gauge fields.

The structure of this paper is as follows. In the next section we
present the basic field equations as well as the Lagrangian of two
types of nonlinear electrodynamic fields. We will also solve the
equations for the rotating black string spacetime and study the
physical properties of the solutions. In Sec. III, we calculate
the conserved and thermodynamic quantities of the black string
solutions and verify the first law of thermodynamics. We finish
our paper with closing remarks in the last section.

\section{Basic equations and solutions}

We consider a model of a gravitating electromagnetic field in the presence
of cosmological constant. The Lagrangian for this system is chosen as%
\begin{equation}
\mathcal{L}=R-2\Lambda +L(\mathcal{F}),  \label{Lagrangian}
\end{equation}%
where ${R}$ is the Ricci scalar, $\Lambda $ refers to the cosmological
constant and $L(\mathcal{F})$ is a general Lagrangian of electromagnetic
field in which $\mathcal{F}=F_{\mu \nu }F^{\mu \nu }$ is Maxwell invariant, $%
F_{\mu \nu }=\partial _{\mu }A_{\nu }-\partial _{\nu }A_{\mu }$ and $A_{\mu
} $ is the gauge potential. Here, we assume a kind of rotating metric whose $%
t$ =constant and $r$ =constant boundary has the topology $R\times S^{1}$
\cite{Lem}
\begin{equation}
ds^{2}=-f(r)\left( \Xi dt-ad\phi \right) ^{2}+\frac{r^{2}}{l^{4}}\left(
adt-\Xi l^{2}d\phi \right) ^{2}+\frac{dr^{2}}{f(r)}+\frac{r^{2}}{l^{2}}
dz^{2},  \label{Metric}
\end{equation}
where the functions $f(r)$ should be determined from gravitational field, $%
\Xi =\sqrt{1+a^{2}/l^{2}}$, $a$ is the rotation parameter, $0\leq \phi <2\pi
$ and $-\infty <z<\infty $.

In order to investigate the properties of the electromagnetic
field, one may consider a suitable $L(\mathcal{F})$ and solve the
Maxwell-like field equation. In this paper, we consider two new
classes of nonlinear electromagnetic fields, namely exponential
form of nonlinear electromagnetic (ENE) Lagrangian and logarithmic
form of nonlinear electromagnetic (LNE) Lagrangian, whose
Lagrangians are
\begin{equation}
L(\mathcal{F})=\left\{
\begin{array}{ll}
\beta ^{2}\left( \exp (-\frac{\mathcal{F}}{\beta ^{2}})-1\right) , & \text{%
ENE} \\
-8\beta ^{2}\ln \left( 1+\frac{\mathcal{F}}{8\beta ^{2}}\right) , & \text{LNE%
}%
\end{array}%
\right. ,  \label{L(F)}
\end{equation}%
where $\beta $ is called the nonlinearity parameter. In the limit $\beta
\rightarrow \infty $, the mentioned $L(\mathcal{F})$'s reduce to the
Lagrangian of the standard Maxwell field
\begin{equation}
\left. L({\mathcal{F}})\right\vert _{\text{large }\beta }=-{\mathcal{F}}+O({%
\mathcal{F}}^{2})  \label{LBIexpand}
\end{equation}

Considering a strong electromagnetic field in regions near to
point-like charges, Dirac suggested that one may have to use
generalized nonlinear Maxwell theory in those regions
\cite{Dirac64}. Similar behavior may have occurred in the vicinity
of neutron stars and black objects and so it is expected to
consider nonlinear electromagnetic fields with an astrophysical
motive \cite{AstroNON}. In addition, within the framework of
quantum electrodynamics, it was shown that quantum corrections
lead to nonlinear properties of vacuum which affect the photon
propagation \cite{H-E,Schwinger,Stehle,Delphenich}.

Although in the context of nonlinear electrodynamics, BI theory is
a specific model, the recent interest in the nonlinear
electrodynamics theories is mainly due to their emergence in the
context of the low-energy limit of heterotic string theory, where
a quartic correction of Maxwell field strength appear \cite{Kats}.
In other words, it was shown that all order loop corrections to
gravity may be added up as a Born-Infeld type Lagrangian
\cite{BIString,Frad,Fradkin}. Any nonlinear electrodynamics that
satisfies the weak field limit (\ref{LBIexpand}) is said to be of
the Born-Infeld type \cite{BItype}.

For completeness, we should note that working in the context of
AdS/CFT correspondence, it is worth investigating the effects of
nonlinear electrodynamic fields on the dynamics of the strongly
coupled dual theory \cite{CaiSun}.

Motivated by the recent results mentioned above, we consider the
mentioned Born-Infeld type theory and investigate their
properties.

The equation of motion for the gauge field can be written as
\begin{equation}
\partial _{\mu }\left( \sqrt{-g}L_{\mathcal{F}}F^{\mu \nu }\right) =0,
\label{MaxEq}
\end{equation}%
where $L_{\mathcal{F}}=\frac{dL(\mathcal{F})}{d\mathcal{F}}$.
Considering Eq. (\ref{Metric}) with Eq. (\ref{MaxEq}), we find
that the consistent gauge potential is
\begin{equation}
A_{\mu }=h(r)\left( \Xi \delta _{\mu }^{0}-a\delta _{\mu }^{\phi }\right) ,
\label{Amu}
\end{equation}%
in which the radial function $h(r)$ can be written as
\begin{equation}
h(r)=\left\{
\begin{array}{ll}
\frac{-\beta r\sqrt{L_{W}}}{2}\left[ 1+\frac{L_{W}}{5}\digamma {\left( \left[
1\right] ,\left[ \frac{9}{4}\right] ,\frac{L_{W}}{4}\right) }\right] , &
\text{ENE}\vspace{0.1cm} \\
\frac{-2q}{3r}\left[ 2\digamma {\left( \left[ \frac{1}{4},\frac{1}{2}\right]
,\left[ \frac{5}{4}\right] ,1-\Gamma ^{2}\right) -}\frac{1}{\left( 1+\Gamma
\right) }\right] , & \text{LNE}%
\end{array}%
\right. ,  \label{h(r)}
\end{equation}%
where $q$ is an integration constant which is related to the electric charge
of the black string, $L_{W}=LambertW\left( \frac{4q^{2}}{\beta ^{2}r^{4}}%
\right) $ which satisfies $LambertW(x)\exp \left[ LambertW(x)\right] =x$, $%
\digamma ([a],[b],z)$ is hypergeometric function and $\Gamma =\sqrt{1+\frac{%
q^{2}}{r^{4}\beta ^{2}}}$ (for more details, see \cite{Lambert}).

\begin{figure}[tbp]
$%
\begin{array}{cc}
\epsfxsize=8cm \epsffile{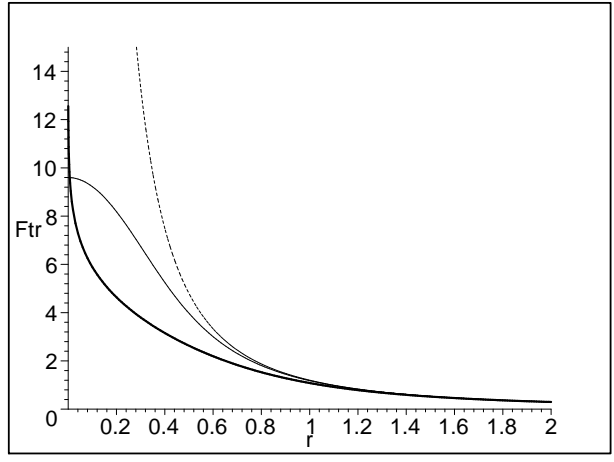} &
\end{array}
$%
\caption{The behavior of $F_{tr}$ versus $r$ for $q=1$, $\protect\beta=4$
and $\Xi=1.2$. ENE (Bold line), LNE (solid line), and linear Maxwell field
(dotted line).}
\label{E}
\end{figure}

Using Eq. (\ref{Amu}), we find that the only nonzero components of
the electromagnetic field tensor are $F_{tr}$ and $F_{\phi r}$,
\begin{equation}
F_{\phi r}=-\frac{a}{\Xi }F_{tr},\;{~\;~}F_{tr}=\frac{\Xi q}{r^{2}}\times
\left\{
\begin{array}{ll}
e^{-\frac{L_{W}}{2}}, & \text{ENE} \\
\frac{2}{\Gamma +1}, & \text{LNE}%
\end{array}%
\right. .  \label{El}
\end{equation}%
Expanding $F_{tr}$ for large value of $r$, we arrive at%
\begin{equation}
F_{tr}=\frac{\Xi q}{r^{2}}+-\frac{\xi \Xi q^{3}}{4\beta ^{2}r^{6}}+O\left(
\frac{1}{r^{10}}\right) ,  \label{Ftrexpand}
\end{equation}
where $\xi =1,8$ for LNE and ENE branches, respectively. We find
that for large distances, the first term in Eqs. (\ref{Ftrexpand})
dominates and the electric field of Maxwell theory is recovered.
We have plotted $F_{tr}$ as a function of $r$ in Fig. \ref{E}.
From this figure it can be seen that for
large values of $r$ the electric fields vanish as one expected. Besides, as $%
r\rightarrow \infty $, both ENE and LNE behave like the linear
Maxwell field. This implies that the nonlinearity of these fields
make sense only near the origin. It is worth mentioning that, in
contrast to the standard Maxwell and ENE fields, the electric
field of LNE is finite at the origin. We should note that the
divergency of the electric field of ENE is much slower than the
divergency of the standard Maxwell field at $r=0$.

Now, we are in a position to discuss the geometric view of
spacetime. To do this, we should first obtain the metric function
$f(r)$. Gravitational field equation of Einstein--$\Lambda
$--nonlinear electromagnetic theory may be written as
\cite{HendiJHEP}
\begin{equation}
R_{\mu \nu }-\frac{1}{2}g_{\mu \nu }\left( R-2\Lambda \right) =\frac{1}{2}%
g_{\mu \nu }L(\mathcal{F})-2L_{\mathcal{F}}F_{\mu \lambda }F_{\nu
}^{\;\lambda }.  \label{GravEq}
\end{equation}%
In order to obtain $f(r)$, we should consider the field Eq. (\ref%
{GravEq}) with Eqs. (\ref{Metric}), (\ref{L(F)}) and (\ref{El}). It is easy
to show that the $rr$ component of Eq. (\ref{GravEq}) can be written in the
following form%
\begin{equation}
rf^{\prime }(r)+f(r)+\Lambda r^{2}+\beta ^{2}r^{2}g(r)=0,  \label{rrEq}
\end{equation}%
where
\begin{equation}
g(r)=\left\{
\begin{array}{ll}
\frac{\exp (8\Psi )}{2}\left[ 1-\left( 1-16\Psi \right) \right] , & \text{ENE%
}\vspace{0.2cm} \\
4\ln \left( 1-\Psi \right) +\frac{8\Psi }{1-\Psi }, & \text{LNE}%
\end{array}%
\right. ,  \label{g(r)Eq}
\end{equation}%
and $\Psi =\left( \frac{F_{tr}}{2\Xi \beta }\right) ^{2}$. After some
cumbersome calculations, the solutions of Eq. (\ref{rrEq}) can be obtained as%
\begin{equation}
f(r)=\frac{2m}{r}-\frac{\Lambda r^{2}}{3}+\beta W(r),  \label{f(r)}
\end{equation}%
with%
\[
W(r)=\left\{
\begin{array}{ll}
\frac{q}{3\sqrt{L_{W}}}\left[ 1+L_{W}+\frac{4}{5}L_{W}^{2}\digamma {\left(
[1],[\frac{9}{4}],\frac{L_{W}}{4}\right) }\right] , & \text{ENE}\vspace{0.2cm%
} \\
\frac{8q^{2}\digamma {\left( [\frac{1}{2},\frac{1}{4}],[\frac{5}{4}%
],1-\Gamma ^{2}\right) }}{3\beta r^{2}}-\frac{4\beta r^{2}\left( \Gamma -\ln
[\frac{(\Gamma ^{2}-1)}{2e^{-7/3}}]\right) }{3}-\frac{4\beta \int r^{2}\ln
\left( \Gamma -1\right) dr}{r}, & \text{LNE}%
\end{array}%
\right. ,
\]%
where $m$ is the integration constant which is the total mass of spacetime, $%
\digamma ([a],[b],z)$ is hypergeometric function and the last term of LNE
can be calculated as%
\begin{eqnarray}
\int r^{2}\ln \left( \Gamma -1\right) dr &=&-\frac{q^{3/2}(\Gamma -1)^{1/4}}{%
2^{3/4}\beta ^{3/2}}\left[ \frac{14}{3}\digamma {\left( \left[ \frac{1}{4},%
\frac{1}{4},\frac{11}{4}\right] ,\left[ \frac{5}{4},\frac{5}{4}\right] ,%
\frac{1-\Gamma }{2}\right) }-\right.  \nonumber \\
&&\left. \frac{14}{25}(\Gamma -1)\digamma {\left( \left[ \frac{5}{4},\frac{5%
}{4},\frac{11}{4}\right] ,\left[ \frac{9}{4},\frac{9}{4}\right] ,\frac{%
1-\Gamma }{2}\right) -}\right.  \nonumber \\
&&\left. \frac{\left[ 4+3\ln (\Gamma -1)\right] }{9(\Gamma -1)}\digamma {%
\left( \left[ \frac{-3}{4},\frac{7}{4}\right] ,\left[ \frac{1}{4}\right] ,%
\frac{1-\Gamma }{2}\right) }+\right.  \nonumber \\
&&\left. \left[ -4+\ln (\Gamma -1)\right] \digamma {\left( \left[ \frac{1}{4}%
,\frac{7}{4}\right] ,\left[ \frac{5}{4}\right] ,\frac{1-\Gamma }{2}\right) }%
\right] .  \label{integral}
\end{eqnarray}

We should note that obtained solutions given by Eq. (\ref{f(r)})
satisfy all the components of the field equations (\ref{GravEq}).

\subsection*{Properties of the solutions}

Here we are going to study the physical properties of the solutions as well
as the asymptotic behavior of the spacetime. Expanding the metric functions
for large $r$, we have%
\begin{equation}
f(r)=\frac{2m}{r}-\frac{\Lambda r^{2}}{3}+\frac{q^{2}}{r^{2}}-\frac{\xi q^{4}%
}{40r^{6}\beta ^{2}}+O\left( \frac{1}{r^{10}}\right) ,  \label{f(r)Expand}
\end{equation}%
where for $\beta \rightarrow \infty $, one can recover the rotating black
string solutions in Einstein-Maxwell gravity \cite{Lem}. Next, we calculate
the curvature scalars of this spacetime. It is easy to show that the Ricci
scalar and the Kretschmann invariant of the spacetime are
\begin{eqnarray}
R &=&-f^{\prime \prime }(r)-\frac{4f^{\prime }(r)}{r}-\frac{2f(r)}{r^{2}},
\label{R} \\
R_{\mu \nu \rho \sigma }R^{\mu \nu \rho \sigma } &=&f^{\prime \prime
2}(r)+\left( \frac{2f^{\prime }(r)}{r}\right) ^{2}+\left( \frac{2f(r)}{r^{2}}%
\right) ^{2}.  \label{RR}
\end{eqnarray}%
where the prime denotes derivative with respect to $r$. Since other
curvature invariants of the spacetime such as the Ricci square are only the
functions of $f^{\prime \prime }$, $f^{\prime }/r$ and $f/r^{2}$, thus we
only consider the Ricci scalar and the Kretschmann invariant. Substituting
the metric functions (\ref{f(r)}) in (\ref{R}) and (\ref{RR}), we find
\begin{eqnarray}
\lim_{r\longrightarrow 0^{+}}R &=&\infty ,  \label{Rorigin} \\
\lim_{r\longrightarrow 0^{+}}R_{\mu \nu \rho \sigma }R^{\mu \nu \rho \sigma
} &=&\infty .  \label{RRorigin}
\end{eqnarray}%
This indicates that we have an essential singularity located at $r=0$. On
the other side, we can expand Eqs. (\ref{R}) and (\ref{RR}) for large $r$
and keep the first order nonlinear correction term%
\begin{eqnarray}
\lim_{r\longrightarrow \infty }R &=&4\Lambda +\frac{\xi Q^{4}}{2\beta
^{2}r^{8}}+O\left( \frac{1}{r^{12}}\right) ,  \label{Rinf} \\
\lim_{r\longrightarrow \infty }R_{\mu \nu \rho \sigma }R^{\mu \nu \rho
\sigma } &=&\frac{8}{3}\Lambda ^{2}+\frac{48m^{2}}{r^{6}}-\frac{96mQ^{2}}{%
r^{7}}+\frac{56Q^{4}}{r^{8}}-\frac{2\xi Q^{4}}{\beta ^{2}l^{2}r^{8}}+O\left(
\frac{1}{r^{11}}\right) ,  \label{RRinf}
\end{eqnarray}%
where we conclude that the spacetime is asymptotically anti-de Sitter.

\begin{figure}[tbp]
$%
\begin{array}{cc}
\epsfxsize=7cm \epsffile{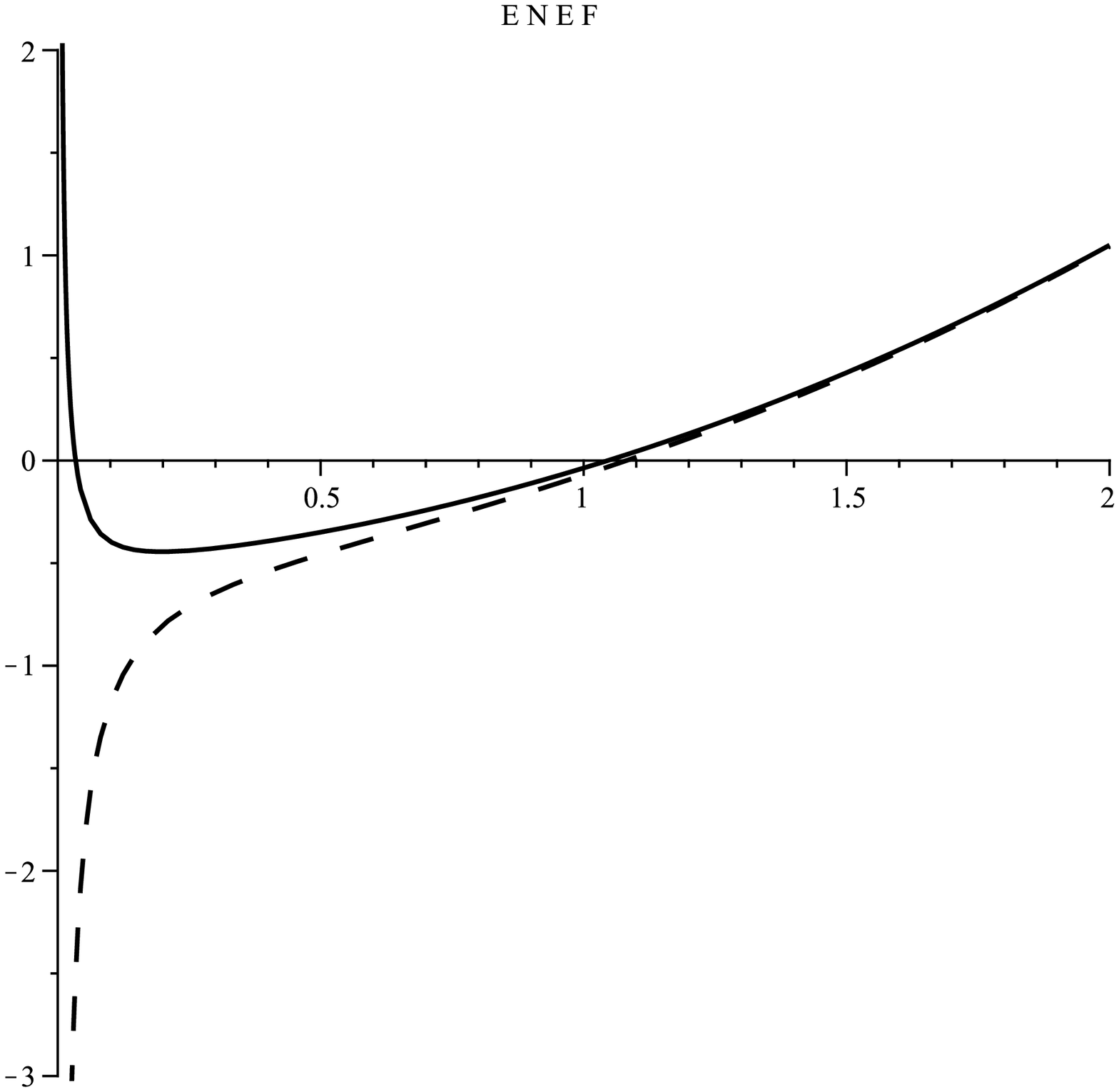} & \epsfxsize=7cm \epsffile{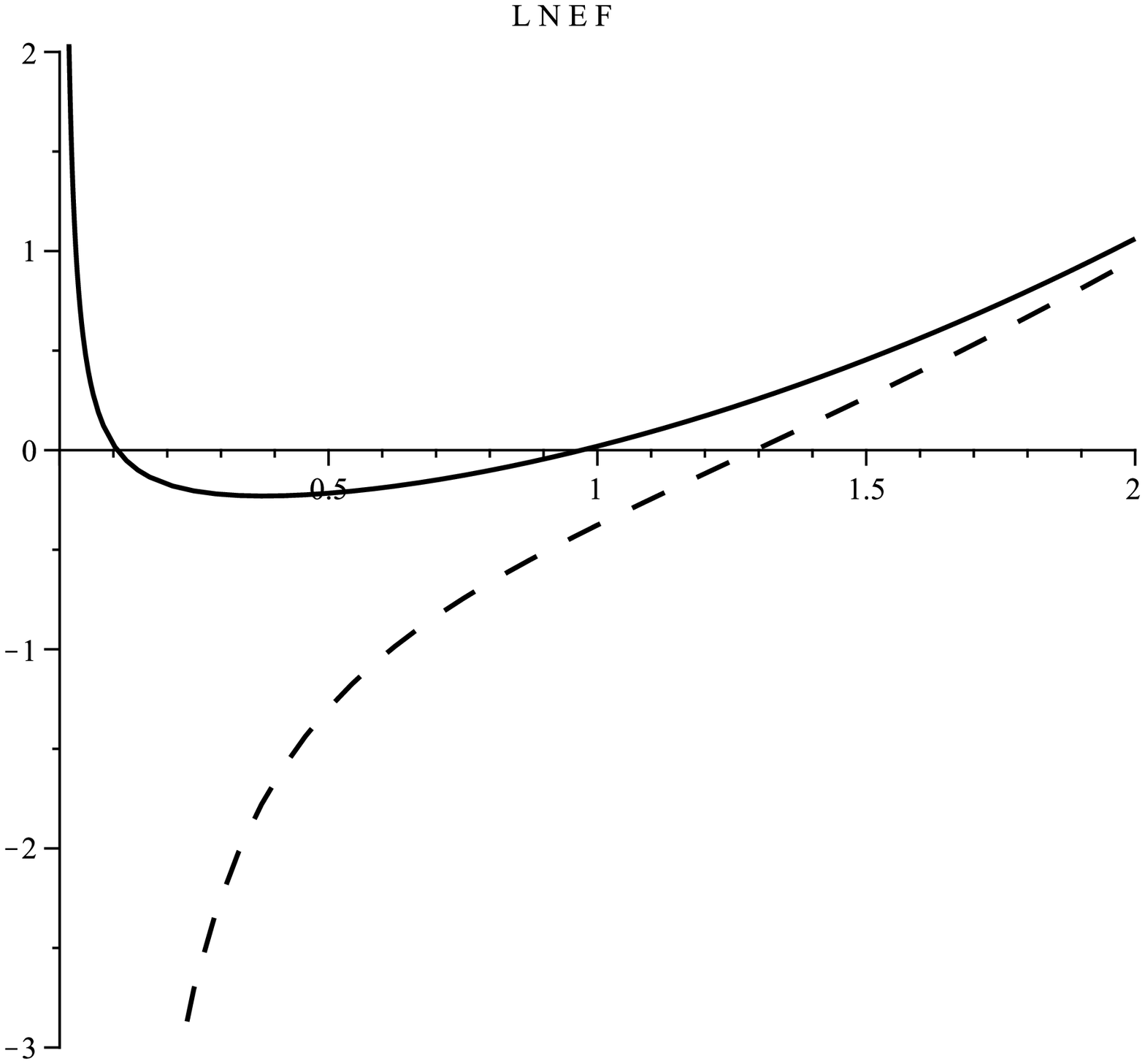}%
\end{array}
$%
\caption{$f(r)$ versus $r$ for $\Lambda=-1$, $q=1$, $m=1$, and $\protect%
\beta > \protect\beta_{c}$ (solid line) and $\protect\beta <\protect\beta%
_{c} $ (dashed line) in ENE branch (left figure) and LNE branch (right
figure). }
\label{Figf}
\end{figure}

Furthermore, such as black string with linear Maxwell source, one
expects to obtain a black string with an outer and an inner
horizon, an extreme black string or a naked singularity. We should
note that for the obtained black string with nonlinear source, a
new interesting situation appears. This new situation appears for
small values of the nonlinearity parameter, in which the black
string has one non-extreme horizon as it happens for Schwarzschild
solution (uncharged solution). In other words, we find that there
is a critical nonlinearity parameter $\beta _{c}$ in which for
$\beta <\beta _{c}$ the metric function may be negative near the
origin. This means that the singularity is timelike for $\beta
>\beta _{c}$, but for $\beta <\beta _{c}$ it is spacelike (see
Fig. \ref{Figf} for more details). Although we cannot obtain
$\beta _{c}$ analytically, we find that it is a function of other
parameters. We obtain $\beta _{c}$, numerically, from the
following
conditions%
\[
\lim_{r\longrightarrow 0^{+}}f(r)\longrightarrow \left\{
\begin{array}{ll}
+\infty , & \beta >\beta _{c} \\
-\infty , & \beta <\beta _{c}%
\end{array}%
\right. .
\]
For more clarifications, we give a numerical method for obtaining $\beta
_{c} $. At first we should fix metric parameters $m$, $q$ and $\Lambda $,
and we can check that for large $\beta $ the metric function is positive
near the origin. Then, we reduce $\beta $ until the sign of the metric
function switch to negative. One can use this method to obtain $\beta _{c}$
with ideal accuracy. Numerical calculations show that $\beta _{c}$ change
when we alter at least one of the metric parameters $m$, $q$ and $\Lambda$
(see table A for more details).

\begin{center}
\begin{tabular}{ccc}
\hline\hline
$f_{ENE}(r)\longrightarrow $ & $-\infty $ & ~~ $+\infty $ \\ \hline
$\beta =$ & ~~ $0.171167$ & ~~ $0.171168$ \\ \hline\hline
\end{tabular}
\\[0pt]
\vspace{0.5cm}
\begin{tabular}{ccc}
\hline\hline
$f_{LNE}(r)\longrightarrow $ & ~~ $-\infty $ & ~~ $+\infty $ \\ \hline
$\beta =$ & ~~ $0.092042$ & ~~ $0.092043$ \\ \hline\hline
\end{tabular}
\\[0pt]
\vspace{0.3cm} Table A: $f(r)$ for $m=1$, $q=1$, $\Lambda =-1$, $%
r\longrightarrow 0^{+}$ and \\[0pt]
$\beta _{c}\approx 0.171$ and $0.092$ for ENE and LNE branches, respectively.
\end{center}

\section{Conserved quantities and the first law of thermodynamics \label{Therm}}

At the first step, we use the definition of quasilocal energy \cite%
{BY,Kraus,Mal} to compute the conserved charges of our solutions. Following
the counterterm method, the divergence-free boundary stress-tensor can be
written as
\begin{equation}
T^{ab}=\frac{\Theta ^{ab}-\left( \Theta +\frac{2}{l}\right) \gamma ^{ab}}{%
8\pi },  \label{Stres}
\end{equation}%
where the last term comes from counterterm procedure. Considering a Killing
vector field $\mathcal{\xi }$ on the boundary $\mathcal{B}$, it is known
that its quasilocal conserved quantity may be obtain from the following
relation%
\begin{equation}
Q(\mathcal{\xi )}=\int_{\mathcal{B}}d^{2}x\sqrt{\sigma }T_{ab}n^{a}\mathcal{%
\ \xi }^{b},  \label{TotalCharge}
\end{equation}%
where $\sigma $ is the determinant of the Arnowitt-Deser-Misner form of
boundary metric $\sigma _{ij}$ and $n^{a}$ is the unit normal vector on the
boundary $\mathcal{B}$. It is easy to find that boundary $\mathcal{B}$ has
two killing vector fields. They are timelike ($\partial /\partial t$) and
rotational ($\partial /\partial \varphi $) Killing vector fields, in which
their corresponding conserved charges are the quasilocal mass and angular
momentum. One can find that the mass and angular momentum per unit length of
the string when the boundary $\mathcal{B}$ goes to infinity can be
calculated as%
\begin{equation}
{M}=\frac{1}{16\pi l}\left( 3\Xi ^{2}-1\right) m,  \label{M}
\end{equation}%
\begin{equation}
{J}=\frac{3}{16\pi l}\Xi ma.  \label{J}
\end{equation}%
As one can find the angular momentum per unit length vanishes for $a=0$ ($%
\Xi =1$) and therefore, $a$ is the rotational parameter,
correctly.

The second step is devoted to calculation of the thermodynamic
quantities. It was known that the universal area law of the
entropy can apply to all types of black objects in Einstein
gravity \cite{Bekenstein,hunt}. Therefore, the entropy per unit
length of the black string is
\begin{equation}
{S}=\frac{\Xi r_{+}^{2}}{4l}.  \label{S}
\end{equation}%
In order to obtain angular velocity $\Omega $ and Hawking temperature of the
black string at the event horizon, we use the method of analytic
continuation of the metric. One can obtain the Euclidean section of the
metric by use of a transformation ($t\rightarrow i\tau $ and $a\rightarrow
ia $) and regularity at $r=r_{+}$ requires that $\tau $ and $\phi $\ should,
respectively, identify with $\tau +\beta _{+}$ and $\phi +i\Omega \beta _{+}$%
, where $\beta _{+}$ is the inverse of the Hawking temperature. So, it is
easy to find that%
\begin{equation}
\Omega =\frac{a}{\Xi l^{2}},  \label{Omega}
\end{equation}%
and%
\begin{equation}
T_{+}=-\frac{\Lambda r_{+}}{4\pi }+\left\{
\begin{array}{ll}
\frac{\beta q\left( 1-L_{W_{+}}\right) }{4\pi r_{+}\sqrt{L_{W_{+}}}}-\frac{%
\beta ^{2}r_{+}}{8\pi }, & \text{ ENE}\vspace{0.2cm} \\
-\frac{q^{2}\left( \Gamma _{+}-2\right) }{\pi r_{+}^{3}\Gamma _{+}\left(
\Gamma _{+}-1\right) }+\frac{\beta ^{2}r_{+}\left( \ln (\frac{\Gamma
_{+}^{2}-1}{2})-\frac{2}{\Gamma _{+}}\right) }{\pi }, & \text{LNE}%
\end{array}%
\right. ,  \label{T}
\end{equation}%
where $\Gamma _{+}=\sqrt{1+\frac{q^{2}}{r_{+}^{4}\beta ^{2}}}$ and $%
L_{W_{+}}=LambertW\left( \frac{4q^{2}}{\beta ^{2}r_{+}^{4}}\right) $.

In order to examine the first law of thermodynamics, we should calculate the
electric charge and potential of the black string. We should use the Gauss'
law and calculate the flux of the electromagnetic field at infinity to
obtain the electric charge per unit length of black string
\begin{equation}
{Q}=\frac{\Xi q}{4\pi l}.  \label{Q}
\end{equation}%
In addition, the electric potential $U$, measured at infinity with respect
to the event horizon $r_{+}$, is defined by \cite{Cvetic}
\begin{equation}
U=A_{\mu }\chi ^{\mu }\left\vert _{r\rightarrow \infty }-A_{\mu }\chi ^{\mu
}\right\vert _{r=r_{+}},  \label{U1}
\end{equation}%
where $\chi =\partial _{t}+\Omega \partial _{\phi }$ is the null generator
of the event horizon. It is easy to show that%
\begin{equation}
U=\frac{1}{\Xi }\times \left\{
\begin{array}{ll}
\frac{\beta r\sqrt{L_{W}}}{2}\left[ 1+\frac{L_{W_{+}}}{5}\digamma {\left( %
\left[ 1\right] ,\left[ \frac{9}{4}\right] ,\frac{L_{W_{+}}}{4}\right) }%
\right] , & \text{ENE}\vspace{0.1cm} \\
\frac{2q}{3r_{+}}\left[ 2\digamma {\left( \left[ \frac{1}{4},\frac{1}{2}%
\right] ,\left[ \frac{5}{4}\right] ,1-\Gamma _{+}^{2}\right) -}\frac{1}{%
\left( 1+\Gamma _{+}\right) }\right] , & \text{LNE}%
\end{array}%
\right. .  \label{U}
\end{equation}

Now, we are in a position to check the first law of black string
thermodynamics. Using the Smarr-type formula, it is straightforward to
calculate the temperature, angular velocity and electric potential in the
following manner
\begin{equation}
T=\left( \frac{\partial M}{\partial S}\right) _{J,Q},\text{ \ \ \ \ \ \ }%
\Omega =\left( \frac{\partial M}{\partial J}\right) _{S,Q},\text{ \ \ \ \ \
\ }U=\left( \frac{\partial M}{\partial Q}\right) _{S,J}.  \label{TOmegaU}
\end{equation}%
One can find that the quantities calculated by Eq. (\ref{TOmegaU}) coincide
with Eqs. (\ref{T}), (\ref{Omega}) and (\ref{U}). Hence we conclude that the
first law of thermodynamics is satisfied in the following form
\begin{equation}
dM=TdS+\Omega d{J}+Ud{Q}.  \label{firstLaw}
\end{equation}
Since the asymptotic behavior of the solutions is the same as
linear Einstein-Maxwell black string, it is expected that the
nonlinearity does not affect on the electrical charge, mass and
angular momentum. Nevertheless, we find that the nonlinearity
affects on the other quantities in which calculated at the
horizon. Although the nonlinear electromagnetic field changes some
of the conserved and thermodynamic quantities, as we expected
\cite{1Law}, these quantities satisfy the first law of black hole
mechanics.

\section{Closing remarks}

Many physical systems in the nature have nonlinear behavior.
Einstein field equations of general relativity is also a system of
nonlinear gravitational field equations which can be applied for
describing various gravitational objects. In order to solve the
gravitational field equations in the presence of a matter field,
one can consider either the linear gauge field such as the Maxwell
electrodynamics or the nonlinear matter field such as the BI
electrodynamics. Static and stationary black object solutions of
these theories have been established and their thermodynamics have
been studied during the past decades (see
\cite{Fern,Tamaki,Dey,Cai1,MHD,EM} and references therein). The
advantages of the nonlinear electrodynamics in comparison to the
Maxwell field is that it avoids the divergences at the origin and
leads to a finite electric field on the point particles.

In this paper as a new step, we considered two types of nonlinear
electrodynamic Lagrangians as source. The first one is called the
logarithmic form and the second one named the exponential form.
Then, we constructed new four-dimensional charged rotating black
string solutions with horizon topology $R\times S^{1}$ coupled to
the nonlinear electrodynamic field. These solutions are
asymptotically anti-de Sitter. If one expand the nonlinear
electromagnetic fields for large $r$, one finds that the
asymptotic behavior of them is similar to the linear Maxwell
field. We also calculated the curvature invariants of the
spacetime and showed that there is indeed a curvature singularity
located at $r=0$. Furthermore, we found that, unlike
Einstein-Maxwell black string solutions, for small values of the
nonlinearity parameter, one can obtain a black string with a
non-extreme horizon. We also calculated the conserved quantities
of the rotating black string such as the mass and the angular
momentum as well as the thermodynamic quantities such as the
temperature and entropy associated with the horizon and checked
that the obtained conserved and thermodynamic quantities satisfy
the first law of black hole thermodynamics.

Finally, it is worthwhile to study the dynamic as well as
thermodynamic stability of the solutions, and investigate the
effects of nonlinearity parameter, $\beta$, on the stability of
the presented solutions. We leave these problems for the future
studies.

\acknowledgments{ We thank Shiraz University Research Council.
This work has been supported financially by Research Institute for
Astronomy \& Astrophysics of Maragha (RIAAM), Iran.}


\begin{thebibliography}{99}
\bibitem{BI} M. Born and L. Infeld, Proc. R. Soc. A \textbf{144}, 425 (1934).

\bibitem{BIString} N. Seiberg and E. Witten, JHEP 09, 032 (1999).

\bibitem{Frad} E. Fradkin and A. Tseytlin, Phys. Lett. B \textbf{163}, 123
(1985);

R. Matsaev, M. Rahmanov and A. Tseytlin, Phys. Lett. B
\textbf{193}, 207 (1987);

E. Bergshoeff, E. Sezgin, C. Pope and P. Townsend, Phys. Lett. B \textbf{188}%
, 70 (1987).

\bibitem{Cal} C. Callan, C. Lovelace, C. Nappi and S. Yost, Nucl. Phys. B
\textbf{308}, 221 (1988);

O. Andreev and A. Tseytlin, Nucl. Phys. B \textbf{311}, 221 (1988);

R. Leigh, Mod. Phys. Lett. A \textbf{04}, 2767 (1989).

\bibitem{Fern} D. L. Wiltshire, Phys. Rev. D \textbf{38}, 2445 (1988);

M. Cataldo and A. Garcia, Phys. Lett. B \textbf{456}, 28 (1999);

S. Fernando and D. Krug, Gen. Rel. Grav. \textbf{35}, 129 (2003).

\bibitem{Tamaki} Tamaki, JCAP \textbf{0405}, 004 (2004);

M. Aiello, R. Ferraro and G. Giribet, Phys. Rev. D \textbf{70}, 104014
(2004);

\bibitem{Dey} T. K. Dey, Phys. Lett. B \textbf{595}, 484 (2004).

\bibitem{Cai1} R. G. Cai, D. W. Pang and A. Wang, Phys. Rev. D \textbf{70},
124034 (2004).

\bibitem{MHD} M. H. Dehghani and H. R. Rastegar-Sedehi, Phys. Rev. D \textbf{%
74}, 124018 (2006);

M. H. Dehghani and S. H. Hendi, Int. J. Mod. Phys. D \textbf{16}, 1829
(2007);

M. H. Dehghani, N. Alinejadi and S. H. Hendi, Phys. Rev. D \textbf{77},
104025 (2008);

S. H. Hendi, J. Math. Phys. \textbf{49}, 082501 (2008).

\bibitem{Tam} T. Tamaki and T. Torii, Phys. Rev. D \textbf{62}, 061501R
(2000);

G. Clement and D. Gal'tsov, Phys. Rev. D \textbf{62}, 124013 (2000);

R. Yamazaki and D. Ida, Phys. Rev. D \textbf{64}, 024009 (2001);

S. S. Yazadjiev, Phys. Rev. D \textbf{72}, 044006 (2005);

I. Stefanov, S. S. Yazadjiev, M. D. Todorov, Phys. Rev. D \textbf{75},
084036 (2007).

\bibitem{SheyBI} A. Sheykhi, N. Riazi and M. H. Mahzoon, Phys. Rev. D
\textbf{74}, 044025 (2006);

A. Sheykhi, N. Riazi, Phys. Rev. D \textbf{75}, 024021 (2007);

M. H. Dehghani, A. Sheykhi and S. H. Hendi, Phys. Lett. B \textbf{659}, 476
(2008);

M. H Dehghani, S. H. Hendi, A. Sheykhi and H. R. Rastegar-Sedehi, JCAP
\textbf{0702}, 020 (2007);

A. Sheykhi, Phys. Lett. B \textbf{662}, 7 (2008).

\bibitem{Soleng} H. H. Soleng, Phys. Rev. D \textbf{52}, 6178 (1995).

\bibitem{HendiJHEP} S. H. Hendi, JHEP \textbf{03}, 065 (2012).

\bibitem{PMIpaper} M. Hassaine and C. Martinez, Phys. Rev. D \textbf{75},
027502 (2007);

M. Hassaine and C. Martinez, Class. Quantum Gravit. \textbf{25}, 195023
(2008);

S. H. Hendi and H. R. Rastegar-Sedehi, Gen. Relativ. Gravit. \textbf{41},
1355 (2009);

S. H. Hendi, Phys. Lett. B \textbf{677}, 123 (2009);

H. Maeda, M. Hassaine and C. Martinez, Phys. Rev. D \textbf{79}, 044012
(2009);

S. H. Hendi and B. Eslam Panah, Phys. Lett. B \textbf{684}, 77 (2010);

S. H. Hendi, Phys. Lett. B \textbf{690}, 220 (2010);

S. H. Hendi, Prog. Theor. Phys. \textbf{124}, 493 (2010);

S. H. Hendi, Eur. Phys. J. C \textbf{69}, 281 (2010);

S. H. Hendi, Phys. Rev. D \textbf{82}, 064040 (2010).

\bibitem{Oliveira} H. P. de Oliveira, Class. Quantum Gravit. \textbf{11},
1469 (1994).

\bibitem{Soleng2} B. L. Altshuler, Class. Quantum Gravit. \textbf{7}, 189
(1990).

\bibitem{Lem} J. P. S. Lemos, Class. Quantum Gravit. \textbf{12}, 1081
(1995);

J. P. S. Lemos, Phys. Lett. B \textbf{353}, 46 (1995).


\bibitem{Dirac64} P. A. M. Dirac, Lectures on Quantum Mechanics, Yeshiva
University, Belfer Gradulate School of Science, New York (1964).

\bibitem{AstroNON} Z. Bialynicka-Birula and I. Bialynicka-Birula, Phys. Rev.
D \textbf{2}, 2341 (1970).

\bibitem{H-E} W. Heisenberg and H. Euler, Z. Phys. \textbf{98}, 714 (1936).
\emph{Translation by:} W. Korolevski and H. Kleinert, \emph{Consequences of
Dirac's Theory of the Positron}, [physics/0605038];

H. Yajima and T. Tamaki, Phys. Rev. D \textbf{63}, 064007 (2001).

\bibitem{Delphenich} D. H. Delphenich, \emph{Nonlinear electrodynamics and
QED}, [arXiv: hep-th/0309108];

D. H. Delphenich, \emph{Nonlinear optical analogies in quantum
electrodynamics}, [arXiv: hep-th/0610088].

\bibitem{Schwinger} J. Schwinger, Phys. Rev. \textbf{82}, 664 (1951).

\bibitem{Stehle} P. Stehle and P. G. DeBaryshe, Phys. Rev. \textbf{152},
1135 (1966).

\bibitem{Kats} Y. Kats, L. Motl and M. Padi, JHEP \textbf{0712}, 068 (2007);

D. Anninos and G. Pastras, JHEP \textbf{0907}, 030 (2009);

R. G. Cai, Z. Y. Nie and Y. W. Sun, Phys. Rev. D \textbf{78}, 126007 (2008).

\bibitem{Fradkin} A. Tseytlin, Nucl. Phys. B \textbf{276}, 391 (1986);

D.J. Gross and J. H. Sloan, Nucl. Phys. B \textbf{291}, 41 (1987).

\bibitem{BItype} H. P. de Oliveira, Class. Quantum Gravit. \textbf{11}, 1469
(1994).

\bibitem{CaiSun} R. G. Cai and Y. W. Sun, JHEP \textbf{09}, 115 (2008);

X. H. Ge, Y. Matsuo, F. W. Shu, S. J. Sin and T. Tsukioka, JHEP \textbf{10},
009 (2008).


\bibitem{Lambert} M. Abramowitz and I. A. Stegun, \emph{Handbook of
Mathematical Functions}, Dover, New York, (1972);

R. M. Corless, G. H. Gonnet, D.E. G. Hare, D. J. Jeffrey, and D.
E. Knuth, Adv. Comput. Math. \textbf{5}, 329 (1996).

\bibitem{BY} J. D. Brown and J. W. York, Phys. Rev. D \textbf{47}, 1407
(1993).

\bibitem{Kraus} V. Balasubramanian and P. Kraus, Commun. Math. Phys. \textbf{%
208}, 413 (1999).

\bibitem{Mal} J. Maldacena, Adv. Theor. Math. Phys. \textbf{2}, 231 (1998);

E. Witten, Adv. Theor. Math. Phys. \textbf{2}, 253 (1998);

O. Aharony, S. S. Gubser, J. Maldacena, H. Ooguri and Y. Oz, Phys. Rep.
\textbf{323}, 183 (2000).

\bibitem{Bekenstein} J. D. Bekenstein, Lett. Nuovo Cimento \textbf{4}, 737
(1972);

J. D. Bekenstein, Phys. Rev. D \textbf{7}, 2333 (1973);

G. W. Gibbons and S. W. Hawking, Phys. Rev. D \textbf{15}, 2738 (1977).

\bibitem{hunt} C. J. Hunter, Phys. Rev. D \textbf{59}, 024009 (1998);

S. W. Hawking, C. J Hunter and D. N. Page, Phys. Rev. D \textbf{59}, 044033
(1999).

\bibitem{Cvetic} M. Cvetic and S. S. Gubser, JHEP \textbf{04}, 024 (1999);

M. M. Caldarelli, G. Cognola and D. Klemm, Class. Quantum Gravit. \textbf{17}%
, 399 (2000).

\bibitem{1Law} D.A. Rasheed, \textit{Nonlinear electrodynamics: zeroth and
first laws of black hole mechanics}, [arXiv:hep-th/9702087].

\bibitem{EM} F. R. Tangherlini, Nuovo Cimento B \textbf{27}, 636 (1963);

R. C. Myers and M. J. Perry, Ann. of Phys. \textbf{172}, 304 (1986);

S. B. Fadeev, V. D. Ivashchuk and V. N. Melnikov, Chinese Phys. Lett.
\textbf{8}, 439 (1991).
\end{thebibliography}
\end{document}